\documentclass{PoS}

\title{Quark propagators at finite temperature with the clover action}

\ShortTitle{Quark propagators at finite temperature with the clover action}

\author{
   \speaker{Masatoshi Hamada},$^a$ Hiroaki Kouno,$^b$ Atsushi Nakamura,$^c$ 
   Takuya Saito,$^d$ Masanobu Yahiro,$^a$\\
\llap{$^a$}Department of Physics, Kyushu University, Fukuoka 812-8581, Japan\\
\llap{$^b$}Department of Physics, Saga University, Saga 840-8502, Japan\\
\llap{$^c$}RIISE, Hiroshima University, Higashi-Hiroshima 739-8521, Japan\\
\llap{$^b$}RCNP, Osaka University, Ibaraki 567-0047, Japan\\
E-mail: \email{hamada2scp@mbox.nc.kyushu-u.ac.jp},
              \email{kounoh@cc.saga-u.ac.jp},
              \email{nakamura@riise.hiroshima-u.ac.jp},
              \email{tsaito@rcnp.osaka-u.ac.jp}, 
              \email{yahiro2scp@mbox.nc.kyushu-u.ac.jp}}

\abstract{
We study properties of the finite temperature quark propagator 
by using the $SU(3)$ quenched lattice simulation in the Landau gauge
and report numerical results of 
the standard Wilson quark case as well as the improved clover one.
The mass function in the deconfinement phase is different from
that of the confinement phase, especially at low momentum regions. 
}

\FullConference{XXIVth International Symposium on Lattice Field Theory\\
                July 23-28, 2006\\
                Tucson, Arizona, USA}
\usepackage{longtable}
\usepackage{mathrsfs}
\usepackage{amsmath,amssymb}
\usepackage{bm}
\usepackage{graphicx}

\begin{document}

\section{Introduction}
Quark propagator is an essential ingredient of QCD and 
should bring us important information on the dynamics of 
the quark-gluon plasma (QGP).
The lattice QCD calculation is a unique tool to investigate
the temperature dependence of the quark propagator, since 
even after the QGP phase transition,
the quark-gluon interaction is still strong, i.e., the QCD coupling
constant is O(1). 
Therefore, we cannot use perturbative calculations.
The RHIC data suggest that the QGP is almost a perfect fluid
indicating that the interaction is very strong.

We first pay an attention to the quark thermal mass, which
is an effective mass proportional to temperature and a consequence of
the preferred reference frame of the heat bath. 
It is studied by perturbative QCD at the region 
that temperature is much higher than critical temperature. 
The results are reviewed in Ref.\cite{tm}. 

Quark propagator at zero temperature is studied by Bowman {\it et al}. 
using lattice QCD
simulation\cite{lhp}. 
The case of clover fermion action is studied by Skullerud and 
Williams\cite{cquark}.
Yet, the quark propagator at finite temperature has not been studied. 

\section{Quark propagator}
The quark propagator in coordinate space is given by 
\begin{align}
G(n,n^{\prime}) = W^{-1}_{f}(n,n^{\prime}),
\end{align}
where $W_f$ is the fermion matrix,
\begin{align}
S_{f} = \sum_{n,n^{\prime}} \bar{\psi}(n) W_{f}(n,n^{\prime}) \psi(n^{\prime}).
\end{align}
We use the clover fermion action given by \cite{ca}
\begin{align}
 S_{f} = S_{W} + S_{C}.
\end{align}
Here $S_{W}$ is Wilson fermions and $S_{C}$ is the clover term. 

It is Fourier transformed to momentum space to calculate the quark mass function and 
the quark thermal mass:
\begin{align}
G(p) = \sum_{n} e^{ip\cdot n}G(n),
\end{align}
where $p_{\mu}$ is the discrete lattice momentum given by
\begin{align}
p_{0} &= \frac{2\pi}{N_{0}}\times \left(n_{0} + \frac{1}{2}\right), \\
p_{i} &= \frac{2\pi}{N_{i}}\times n_{i}.
\end{align}
where anti-periodic (periodic) boundary condition 
at temporal (space) direction is imposed.

At finite temperature, where $O(4)$ invariance is broken and $O(3)$ 
is invariant,
the quark propagator is given by three parameters.
Consequently, the general form of the quark propagator at finite temperature is
\begin{align}
G(\omega_{n},p) &= \frac{1}{i\gamma_{0}\omega_{n}A(\omega_{n},p) 
+ i\gamma_{i}p_{i}B(\omega_{n},p) + C(\omega_{n},p)} \notag\\
&=  \frac{1}{A(\omega_{n},p)}\frac{1}{i\gamma_{0}\omega_{0}
+ i\gamma_{i}p_{i}B^{\prime}(\omega_{n},p) + C^{\prime}(\omega_{n},p)}.
\end{align}
We define the mass function as \cite{cquark}
\begin{align}
M(\omega_{n}, p) = C^{\prime}(\omega_{n}, p) - C^{\prime}_{free}(\omega_{n}, p).
\end{align}
The thermal mass is independent of momentum. The mass is then defined as
\begin{align}
M_{thermal} = M(0). \label{eq:thermal_mass}
\end{align}
The quark thermal mass at high temperature is given by perturbative calculation \cite{tm},
\begin{align}
m_{thermal}^{2} = \frac{1}{8}g^{2}T^{2}C_{F}. \label{eq:t_mass_pqcd}
\end{align}
$g$ is the coupling constant and $C_{F}(=4/3)$ is a group factor.

\section{Numerical results}
We have calculated the quark propagator using the Wilson fermion action and the clover fermion action.
The $SU(3)$ quenched lattice simulation in the Landau gauge is performed on 
a $16^{3}\times$8 lattice.

\subsection{Mass function}
\begin{figure}[t]
\begin{center}
 \includegraphics[scale=0.3]{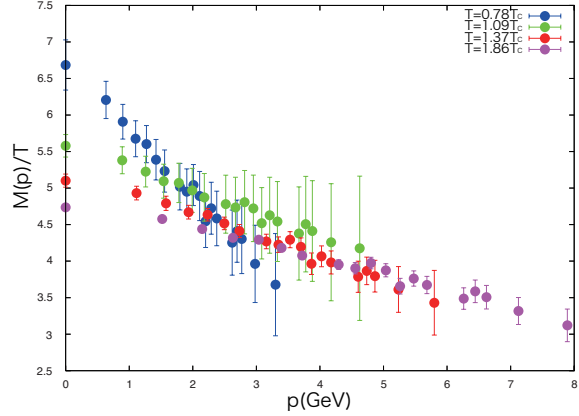}
\end{center}
\caption{Momentum dependence of the mass function using Wilson fermion. 
Blue dots are the case of the confinement phase
($T=0.78T_{c}$). Other three color dots correspond to the deconfinement phase
($T=1.09T_{c}$,$T=1.37T_{c}$,$T=1.86T_{c}$). }
\label{fig:1}
\end{figure}

In Figure \ref{fig:1}, the mass function is plotted using the Wilson fermion action. 
Blue dots correspond to the case of the confinement phase
($T=0.78T_{c}$). Other three color dots are cases of the deconfinement phase
($1.09T_{c}$, $1.37T_{c}$ and $1.86T_{c}$). 
In the high momentum region, the mass function decreases for both the cases.
This behavior represents asymptotic freedom. 
 In the low momentum region, in contrast, mass function of both cases has different behavior
 for the two cases.
The mass function of the confinement phase increases rapidly as momentum decreases,
while that of the deconfinement phase increases slowly
as momentum decreases.

Our first result is that the quark mass function of the deconfinement phase 
is different from that of the confinement phase in which the quark propagator
should not have a physical pole. 

\subsection{Thermal mass}
\begin{figure}[htbp]
\begin{center}
 \includegraphics[scale=0.3]{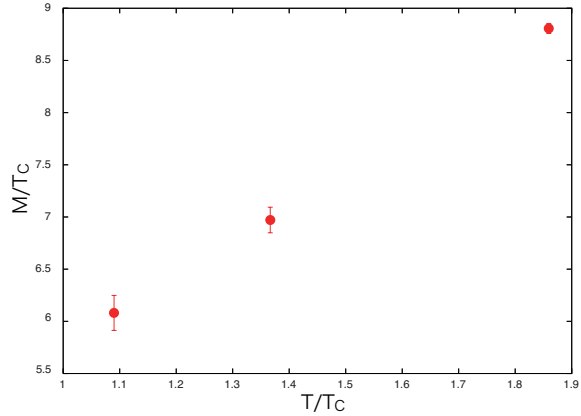}
\end{center}
\caption{Temperature dependence of the quark thermal mass defined by mass function. 
Thermal mass is shown as a linear function of temperature.}
\label{fig:2}
\end{figure}

In Figure \ref{fig:2}, the thermal mass is plotted using the Wilson fermion action. 
It is evaluated from the mass function at zero momentum.
The thermal mass increases linearly as a function of temperature(T).
This is consistent with perturbative calculation (\ref{eq:t_mass_pqcd}).

\subsection{Clover fermion action versus Wilson fermion action}
\begin{figure}[htbp]
\begin{center}
 \includegraphics[scale=0.3]{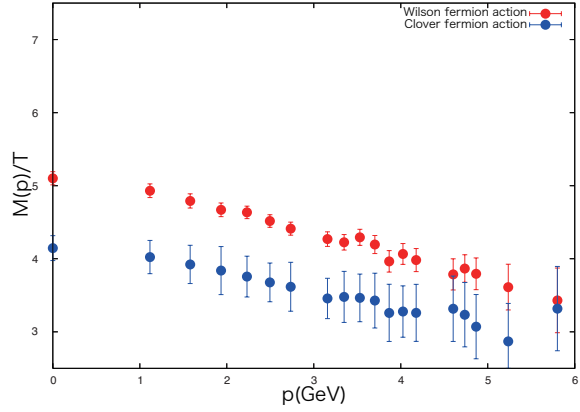}
\end{center}
\caption{Momentum dependence of the mass function. Temperature is $1.37T_{c}$.
Red dots are the case of Wilson fermion action and blue dots are the case of clover fermion action.}
\label{fig:3}
\end{figure}

 In Figure \ref{fig:3}, we compare the mass function based on the clover fermion action with 
that on the Wilson fermion action
in order to study finite lattice space effect at $T=1.37T_{c}$.
Red dots stand for the case of Wilson fermion action and blue dots are the case of clover fermion 
action.
The mass function has qualitatively same behavior for the two cases.
However, the mass function based on the Wilson fermion action overestimates
that on the clover fermion action by about 20$\%$.

\section{Summary}
We study the finite temperature quark propagator 
by using the $SU(3)$ quenched lattice simulation in the Landau gauge.
We have focused an attention on the quark mass function.
First, the quark mass function of confinement phase is different from 
that of deconfinement phase at low momentum region. 
It suggest that in QGP quark behaves on a single particle.
Second, the quark thermal mass is shown as a linear function of temperature. 
Third, the mass function based on clover fermion action
is qualitatively same behavior with and quantitatively different behavior from
that on Wilson fermion action.

\section{Acknowledgements}
The simulation has been done on a supercomputer (NEC SX-5) at 
Research Center for Nuclear Physics, Osaka University.
This work is partially supported by Grant-in-Aid for Scientific
Research by Monbu-Kagakusho, No. 13135216, 17340080 and 18540280.

\end{document}